\documentclass[12pt,aps,pra,reprint,superscriptaddress]{revtex4}

\usepackage{amsmath}
\usepackage{amsfonts}
\usepackage{amssymb}
\usepackage{multirow}
\usepackage{verbatim}
\usepackage{alltt}
\usepackage{moreverb}
\usepackage{graphicx,color,graphics}
\usepackage{hyperref}
\usepackage{setspace}

\begin{document}
\singlespacing

\title{The statistical strength of experiments to reject local realism with photon pairs and inefficient detectors}

\author{Yanbao Zhang}
 \affiliation{Department of Physics, University of Colorado at Boulder, Colorado, 80309,USA}
 \affiliation{Mathematical and Computational Sciences Division, National Institute of Standards and Technology, Boulder, Colorado, 80305,USA}
\author{Emanuel Knill}
 \affiliation{Mathematical and Computational Sciences Division, National Institute of Standards and Technology, Boulder, Colorado, 80305,USA}
\author{Scott Glancy}
 \affiliation{Mathematical and Computational Sciences Division, National Institute of Standards and Technology, Boulder, Colorado, 80305,USA}



\date{\today}

\begin{abstract}
Because of the fundamental importance of Bell's theorem, a
loophole-free demonstration of a violation of local realism (LR) is
highly desirable. Here, we study violations of LR involving photon
pairs. We quantify the experimental evidence against LR by using
measures of statistical strength related to the Kullback-Leibler (KL)
divergence, as suggested by van Dam \emph{et al.} [W. van Dam,  R. Gill and 
P. Grunwald, IEEE Trans. Inf. Theory. \textbf{51}, 2812 (2005)]. Specifically, 
we analyze a test of LR with entangled states created from two independent polarized
photons passing through a polarizing beam splitter. We numerically
study the detection efficiency required to achieve a specified
statistical strength for the rejection of LR depending on whether
photon counters or detectors are used. Based on our results, we find
that a test of LR free of the detection loophole requires photon
counters with efficiencies of at least $89.71\,\%$, or photon detectors
with efficiencies of at least $91.11\,\%$. For comparison, we also perform
this analysis with ideal unbalanced Bell states, which are known to
allow rejection of LR with detector efficiencies above $2/3$.
\end{abstract}



\maketitle

\section{INTRODUCTION}

In 1964, J. Bell first showed that the predictions of quantum
mechanics contradict those of any theory based on local hidden
variables~\cite{Bell}. Such theories are called ``local realistic
theories'', and the principle they are based on is called ``local
realism'' (LR). To disprove local realistic theories, Bell and
others constructed the Bell inequalities. These inequalities are
satisfied by all the predictions of local realistic theories, but are
violated by some predictions of quantum mechanics (see
Refs.~\cite{Peres,Werner,Horodecki} for reviews). The most famous and
easiest to test is the Clauser-Horne-Shimony-Holt (CHSH) inequality~\cite{Clauser}. To test this
inequality, each of two parties---Alice and Bob---receives one particle
from a common source. Each of them performs one of two possible
measurements randomly and independently on their own particle and
records the outcome. This procedure is repeated a large number of
times. At the end, Alice and Bob test the CHSH inequality by analyzing
their joint measurement outcomes. A test of LR showing violation was
first realized by Freedman and Clauser in 1972~\cite{Freedman}.
Since then, many such tests have been performed, which show that
quantum mechanics contradicts LR. For a review, see
Ref.~\cite{Genovese}. However, all tests of LR thus far have required
supplementary assumptions. (We assume without saying that tests of LR
are intended to show violations of LR.) These additional assumptions
introduce two loopholes: the detection loophole~\cite{Pearle} and the
locality loophole~\cite{Bell2}.

The detection loophole is introduced when correlated pairs are
detected with imperfect detectors~\cite{Pearle}. If the detection
efficiencies are sufficiently low, then it is possible for the
subensemble of detected pairs to give results violating LR, even
though the entire ensemble is consistent with LR. To close this
loophole, highly efficient detectors are required, as shown in
Refs.~\cite{Garg,Eberhard,LarssonSemitecolos,Cabello,Brunner}. The
locality loophole arises when there is the possibility of a causal
connection between the event where the measurement setting is chosen
at one site and the event where the measurement outcome is recorded on the
other~\cite{Bell2}. Closing this loophole requires first that the
choices of local measurements should be made randomly and
independently, and second that the distance between different parts of
the experiment should be large enough to prevent light-speed
communication between one observer's measurement choice and the result
of the other observer's measurement.

To date, no single experiment has closed both the detection loophole
and the locality loophole. An experiment carried out on trapped ions
closed the detection loophole~\cite{Rowe}, but the ions were only a
few micrometers apart, so this experiment did not close the locality
loophole. There have been photonic experiments addressing the locality
loophole~\cite{Aspect,Weihs,Tittel}. Yet due to low photon detection
efficiency, photonic experiments have not closed the detection
loophole. A loophole-free test of LR would not only show that some
quantum systems cannot be described by a local realistic theory, but
would also show that a family of quantum communication protocols are
secure even for causal adversaries not limited by the laws of quantum
mechanics~\cite{Barrett, Masanes1, Masanes2}. Hence, it is desirable
to realize an experiment that can demonstrate a loophole-free
violation of LR.

Previous results show that closure of the detection
loophole requires a minimum detection efficiency of $82.85\,\%$
when Bell states are used~\cite{Garg}. With unbalanced Bell states of
the form $\cos(\theta)|00\rangle+\sin(\theta)|11\rangle$, the minimum
detection efficiency approaches $2/3$ as $\theta$ goes to
$0$~\cite{Eberhard}. Recently, a new type of photon counter with high
detection efficiency ($\sim 95\,\%$) was demonstrated~\cite{Lita}, 
making a loophole-free test of LR very promising.

Here, we study the possibility of testing LR with a source of
entangled states created from two independent polarized photons
passing through a polarizing beam splitter. Similar sources are used
in Refs.~\cite{Shih,Ou,Kiess}. We call this source the ``independent
inputs'' source. Although this source does not produce balanced or
unbalanced Bell pairs (see below), it does create some entanglement.
An advantage of this source is that the input photons do not need to
be entangled. The two independent polarized photons can be generated
by spontaneous parametric down-conversion (SPDC) in nonlinear
crystals~\cite{Shih,Ou,Kiess}, or by other single-photon sources being
developed such as atoms, ions, molecules, solid-state quantum dots, or
nitrogen-vacancy centers in diamond~\cite{Lounis,Oxborrow}. The states
of the two photons can be detected by photon counters or photon
detectors. (We use the term ``photon detector'' to refer to detectors
that determine only the presence or absence of photons, not their
number.) Since experimenters can gain more information with photon
counters than with simple photon detectors, we expect that photon
counters make violation of LR more detectable. We also expect that
photon counters can mitigate the influence of the effectively
unentangled part of the state. Our results show that it is possible to
perform a test of LR free of the detection loophole using the
independent inputs source, assuming that the detection efficiency of
photon counters (photon detectors) is at least $89.71\,\%$ (at least
$91.11\,\%$, respectively), showing a small advantage for photon
counters. Furthermore, we numerically quantify the statistical
strength of such a test of LR as a function of the counter or detector
efficiency and state parameters. For comparison, we obtain the same
information for an ideal source of unbalanced Bell states. This makes
it possible to estimate the minimum number of experiments required to
gain reasonable confidence in rejecting LR, as this number is
inversely related to statistical strength.

In Sec.~\ref{sect:expconf}, we briefly describe the experimental scheme
that we analyze. In Sec.~\ref{sect:deficiencies}, we point out the
deficiencies of the most commonly used method for quantifying
violation of LR and summarize the method based on Kullback-Leibler
(KL) divergence proposed in Ref.~\cite{VanDam}. We present our
results in Sec.~\ref{sect:results}. Finally in
Sec.~\ref{sect:conclusion}, we conclude.

\section{EXPERIMENTAL CONFIGURATION}
\label{sect:expconf}

\begin{figure}[htb!]
   \includegraphics[bb=7cm 12.5cm 13cm 20.3cm]{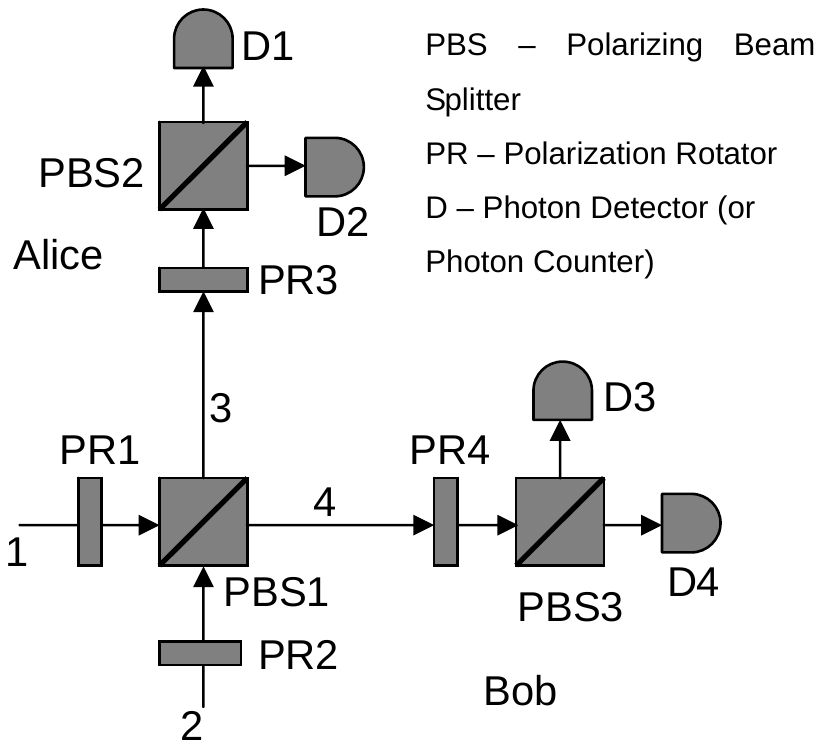}
   \small{
   \caption{Schematic of a test of LR with the independent photons
   source. Two spatially and temporally matched polarized photons are
   inserted at 1 and 2. The polarization rotators PR1 and PR2 are set
   so that photons 1 and 2 are linearly polarized at equal angles when
   they reach the polarizing beam splitter PBS1. After PBS1, the photons
   are in a nonmaximally entangled state [see Eq.~\eqref{real state}]
   and are sent to Alice's and Bob's detector setups. Each detector
   setup uses a PR, a PBS and two detectors. The PR is used to select
   measurement bases by rotating the photon's polarization state.}
   \label{exp. setup}}
\end{figure}

Here we consider a test of LR using pairs of matched polarized
photons. The two photons can be generated by an SPDC
process~\cite{Shih,Ou,Kiess} in the weak-pumping regime,
although single-photon sources could be used~\cite{Lounis,Oxborrow}. Given such
photon pairs, they can be processed as shown in Fig.~\ref{exp. setup}
to produce a state that can violate LR.

Consider a pair of photons arriving in modes 1 and 2 of Fig.~\ref{exp.
setup} in the state
\begin{equation}
|\psi\rangle_{12}=|H\rangle_1 |H\rangle_2,
\end{equation}
where \emph{H} (\emph{V}) denotes horizontal (vertical)
polarization. We set the polarization rotators PR1 and PR2 to
the same angle to produce the state
\begin{equation}
|\psi'\rangle_{12}=(\alpha|H\rangle_1+\beta|V\rangle_1)(\alpha|H\rangle_2+\beta|V\rangle_2),
\end{equation}
where $|\alpha|^2+|\beta|^2=1$. After polarizing beam splitter PBS1,
we get the ``pseudo-Bell'' state
\begin{align}
|\psi_{\text{pB}}\rangle &=\alpha^2|H\rangle_3|H\rangle_4+\beta^2|V\rangle_3|V\rangle_4
         \nonumber \\                                        
           &\qquad +\alpha\beta|H\rangle_3 |V\rangle_3+\alpha\beta|H\rangle_4 |V\rangle_4\label{real state}.
\end{align}

Using these states, we can perform a test of LR. Motivated by the
result of Eberhard~\cite{Eberhard}, we investigate the
possibility of reducing the minimum detection efficiency required to
close the detection loophole in a test of LR by changing the values
of $\alpha$ and $\beta$ in Eq.~\eqref{real state}.

When we set $|\alpha|=|\beta|=1/\sqrt{2}$ in Eq.~\eqref{real state}
and condition on coincidence postselection, we may treat the
pseudo-Bell state as a maximally entangled state, as in the
experiments reported in Refs.~\cite{Shih,Ou,Kiess}. This postselection
process discards events where both photons leave PBS1 in the same
direction, effectively projecting onto a Bell state. However, the
discarded events may create another loophole similar to the detection
loophole for tests of LR~\cite{Kwiat,Caro}. To close this loophole,
the entire pattern of experimental data must be included when
evaluating the terms of a Bell inequality~\cite{Popescu}. Here, we
also use all data without postselection, but instead of obtaining a
violation of a Bell inequality, we quantify the experimental evidence
against all local realistic theories by means of measures derived from
the KL divergence.

\section{DATA ANALYSIS METHOD}
\label{sect:deficiencies}

Contradictions between experimental results and LR are
often shown by the violation of a Bell inequality, such as the CHSH
inequality~\cite{Clauser}
\begin{equation}
E(\hat{A}_1,\hat{B}_1)-E(\hat{A}_1,\hat{B}_2)+E(\hat{A}_2,\hat{B}_1)+E(\hat{A}_2,\hat{B}_2)\leq2,
\end{equation}
where the terms $E(\hat A_a, \hat B_b)$ are correlations between
Alice's and Bob's measurements at settings $\hat A_a$ and $\hat B_b$,
$a,b \in \{1, 2\}$. Following this approach, the departure of an
experiment's results from LR is typically given in terms of
the number of standard deviations separating the experimental value of
the left-hand side of the CHSH inequality from the upper bound of this
inequality, which is $2$. Of course, for any finite set of data, there
is a small probability that a system governed by LR could
also violate the inequality. The standard deviation partially
characterizes the measurement uncertainty due to a finite number of
trials, but it does not consider the probability that a local
realistic system could also violate the inequality. Because such a
system's (non-)violation can have larger standard deviations, the
experimental standard deviation may suggest more confidence in
rejecting LR than justified. To avoid this problem, we
apply a method proposed by van Dam \emph{et al.}~\cite{VanDam}. In this
method, the statistical strength of a test of LR is characterized by
the KL divergence from the experimental statistics to the best
prediction by local realistic theories. The method is justified by the
observation that the confidence at which the experimental data
violate LR is closely related to this KL
divergence~\cite{Bahadur}.

To better understand the approach based on the KL divergence, it is
helpful to analyze tests of LR in terms of a two-player
game. The two players are the quantum experimenter QM and the
theoretician LRT who wants LR to prevail. During the test
of LR, given a source of quantum states, experimenter QM can
randomly change the measurement settings. After a large number $N$ of
trials, QM obtains empirical frequencies $\mathbf{q}$ of measurement
settings and outcomes from the experimental data, which, hopefully,
are consistent with the quantum prediction and violate LR.
At the same time, knowing the state preparation procedure and the
distribution of measurement settings but not the actual settings or
outcomes, LRT can design all kinds of different local realistic
theories, predicting different probability distributions $\mathbf{p}$
for the settings and outcomes. (We are assuming that state preparation
protocols and measurement setting distributions are not changed during
the experiment.) The goal is to make $\mathbf{p}$ as consistent as
possible with the eventually obtained frequencies $\mathbf{q}$. This
requires minimizing a distance between the QM's frequencies
$\mathbf{q}$ and LRT's prediction $\mathbf{p}$. Following the argument
in Ref.~\cite{VanDam}, this distance can be measured by the KL
divergence from $\mathbf{q}$ to $\mathbf{p}$, as defined by
\begin{equation}
D_{\text{KL}}(\mathbf{q}\parallel \mathbf{p})=\sum_{k=1}^K\sum_{l=1}^L q_{kl}\log_{2}\left(\frac{q_{kl}}{p_{kl}}\right)\label{KL divergence},
\end{equation}
where $k$ is the measurement setting index, $K$ is the number of
different measurement settings, $l$ is the measurement outcome index,
and $L$ is the number of different measurement outcomes under each
measurement setting. For example, in the test of the CHSH inequality
using photon pairs maximally entangled in polarization, $k$ denotes
one of the measurement settings $(\hat{A}_1, \hat{B}_1)$, $(\hat{A}_1,
\hat{B}_2)$, $(\hat{A}_2, \hat{B}_1)$, or $(\hat{A}_2, \hat{B}_2)$,
and so $K=2\times 2=4$; $l$ denotes one of the outcomes (\emph{H},
\emph{H}), (\emph{H}, \emph{V}), (\emph{V}, \emph{H}), or (\emph{V},
\emph{V}) (assuming perfect detection) and so $L=2\times 2=4$.

The KL divergence has the property that $D_{\text{KL}}(\mathbf{q}\parallel
\mathbf{p})\geq0$, with equality if and only if $\mathbf{p}
=\mathbf{q}$. Since there are many different local realistic theories,
LRT has the freedom to choose the best one $\mathbf{p}^{(s)}$, namely, 
the one that minimizes the KL divergence. We can then define a
distance from $\mathbf{q}$ to the best local realistic theory
according to
\begin{equation}
D_{\text{KL}}(\mathbf{q}\parallel \mathbf{p}^{(s)})=\min_{\mathbf{p}\in \mathit{P}} D_{\text{KL}}(\mathbf{q}\parallel \mathbf{p})\label{statistical strength},
\end{equation}
where $\mathit{P}$ is the set of local realistic theories. Likewise,
QM also has the freedom to choose different measurement settings and
setting distributions so that the best local realistic theory explains
the experimental data poorly. Hence, the general problem is to
determine the maximum statistical strength $S$ of tests of LR
subject to experimental constraints, which is defined to be
\begin{equation}
S\equiv D_{\text{KL}}\left(\mathbf{q}^{(s)}\parallel \mathbf{p}^{(s)}\right)=\max_{\mathbf{q}\in \mathit{Q}}\min_{\mathbf{p}\in P}D_{\text{KL}}(\mathbf{q}\parallel \mathbf{p})\label{general statistical strength},
\end{equation}
where $\mathbf{q}^{(s)}$ is an optimal quantum strategy maximizing
Eq.~\eqref{statistical strength}, and $\mathit{Q}$ is the set of
accessible quantum strategies. The statistical strength is
asymptotically related to the $p$-value for rejection of LR.
There is a statistical test such that if
$S>0$, then for almost all infinite
sequences of outcomes of independent experiments, 
the  $p$-value after $N$ experiments is bounded by 
\begin{equation} 
p_N=2^{-NS+o(N)},\label{confidence}
\end{equation} 
where $o(N)$ is a data-dependent term that goes to $0$ as
$N\rightarrow\infty$~\cite{Bahadur}. No statistical test can have a
better asymptotic $p$-value. Because $1-p_N$ can be thought of as a
confidence in rejecting LR, the statistical strength $S$ quantifies
the asymptotic rate at which confidence is gained. In particular,
the number of experiments required to have reasonable confidence
in rejecting LR is necessarily greater than $1/S$.

LRT's effort to minimize the KL divergence as in Eq.~\eqref{statistical
strength} is a maximum likelihood estimation problem. Here, we use the
expectation-maximization algorithm in Ref.~\cite{Vardi}. The
general problem of computing the statistical strength $S$ is
nontrivial. To calculate $S$, we maximize Eq.~\eqref{statistical
strength} over measurement settings with standard nonlinear
optimization techniques.

To calculate the statistical strength of a test of LR, we
need to learn how LRT predicts the measurement results given the state
preparation procedure and possible measurement settings. Suppose that for a
bipartite system with $n_A\times n_B$ measurement settings there
are $d_A$ outcomes for each of $n_A$ measurement settings at Alice's
side, and there are $d_B$ outcomes for each of $n_B$ measurement
settings at Bob's side. Then the local realistic description implies
the existence of a single joint probability distribution over a
$d_A^{n_A} \times d_B^{n_B}$-element event space, which we write as
\begin{equation}
P_{\text{LR}}\left(a_1, \ldots, a_{n_A};b_1, \ldots, b_{n_B}|\hat{A}_1, \ldots, \hat{A}_{n_A}; \hat{B}_1, \ldots, \hat{B}_{n_B}\right),
\end{equation}
where $a_1, \ldots, a_{n_A}\in\{1,2,\ldots,d_A\}$, and
$b_1,\ldots,b_{n_B}\in\{1,2,\ldots,d_B\}$, with normalization
\begin{equation}
\sum_{a_1, \ldots, a_{n_A}=1}^{d_A}\hspace{0.5cm}\sum_{b_1, \ldots, b_{n_B}=1}^{d_B}P_{\text{LR}}\left(a_1, \ldots, a_{n_A}; b_1, \ldots, b_{n_B}|\hat{A}_1, \ldots, \hat{A}_{n_A}; \hat{B}_1, \ldots, \hat{B}_{n_B}\right)=1
\end{equation}
Hence, the marginal probability for the measurement outcome ($a_i$;
$b_j$) when settings $\hat{A}_i$ and $\hat{B}_j$ are chosen is given by
\begin{align}
P_{\text{LR}}(a_i; b_j|\hat{A}_i; \hat{B}_j)&=\sum_{a_1, \ldots, a_{i-1},
a_{i+1}, \ldots, a_{n_A}=1}^{d_A}\hspace{0.5cm}\sum_{b_1, \ldots, b_{j-1},
b_{j+1}, \ldots, b_{n_B}=1}^{d_B} \hspace*{4cm}\nonumber \\ &\qquad \hspace*{2cm}P_{\text{LR}}\left(a_1, \ldots,
a_{n_A}; b_1, \ldots, b_{n_B}|\hat{A}_1, \ldots, \hat{A}_{n_A}; \hat{B}_1,
\ldots, \hat{B}_{n_B}\right).
\end{align}
Since the probabilities $P_{\text{LR}}(a_i; b_j|\hat{A}_i; \hat{B}_j)$
are constrained to be marginal distributions, they satisfy nontrivial
relationships. The goal of a test of LR is to choose states and
settings that result in quantum predictions that cannot be obtained as
the marginals of a single local realistic theory for all $i$ and $j$.
The quantum-mechanical prediction of the probability is given by
$P_{\text{qm}}(a_i; b_j|\hat{A}_i; \hat{B}_j)=\text{Tr}(\rho
O(a_i;b_j|\hat{A}_i;\hat{B}_j))$, where $\rho$ is the density matrix
of the quantum state, and $O(a_i; b_j|\hat{A}_i; \hat{B}_j)$ is the
positive operator valued measure (POVM) element corresponding to the
measurement outcome ($a_i$; $b_j$) when Alice and Bob use settings
$\hat{A}_i$ and $\hat{B}_j$, respectively. Given the distributions of
measurement settings chosen by Alice and Bob, the KL divergence
measures the statistical distance of the optimal local realistic
theory from the quantum predictions as in Eq.~\eqref{statistical
strength}.

\section{Results and Discussion}
\label{sect:results}

We consider tests of LR using the independent inputs source for
pseudo-Bell pairs and tests using unbalanced Bell pairs. In both cases,
Alice and Bob use measurement devices like those shown in
Fig.~\ref{exp. setup}. They use either counters or detectors for
photon detection, and they independently and uniformly randomly choose 
one of two measurement settings each, where the settings are determined 
by the polarization rotators. We use Bloch-sphere Euler angles as explained
below to define the measurement settings. We label the measurement
settings $\hat{A}_1$ and $\hat{A}_2$ (Alice) or $\hat{B}_1$ and
$\hat{B}_2$ (Bob) and write the two-photon state coming in at modes 3
and 4 in Fig.~\ref{exp. setup} as $|\psi\rangle_{AB}$. We calculate
the statistical strength $S$ according to Eq.~\eqref{general
statistical strength} by maximizing over the angles of the measurement
settings $\{\hat{A}_1,\hat{A}_2,\hat{B}_1,\hat{B}_2\}$ and minimizing
over the set of local realistic theories $\mathit{P}$, where we fix
the two-photon state $|\psi\rangle_{AB}$ shared by Alice and Bob. The
inner minimization as implemented guarantees convergence to the
optimum $\mathbf{p}^{(s)}$, whereas the outer one obtains a local
optimum. Confidence in global optimality can be obtained by repetition
from many different starting points (which we have done) or more
sophisticated search strategies. A local optimum satisfying $S>0$ is
sufficient for having found a detection-loophole-free test. On the
other hand, finding no solution with $S>0$ is heuristic evidence that
such a test does not exist subject to the constraints of the
experiment. Thus, with this optimization strategy, we can trace the
boundary of the region for which $S>0$ (by searching for where $S$
decreases to $0$) to heuristically determine the minimum detection
efficiency $\eta_{\text{min}}$ and the associated optimal measurement
settings
$\{\hat{A}_{\text{1min}},\hat{A}_{\text{2min}},\hat{B}_{\text{1min}},\hat{B}_{\text{2min}}\}$
needed to perform a test of LR of this type free of the detection
loophole with a given state.

Note that as $S\to 0$, the number of experiments required to gain
confidence close to unity diverges. For a constant rate of gaining
confidence [see the explanation below Eq.~\eqref{confidence}], we set the desired
statistical strength $S=X>0$ and determine the minimum detection
efficiency $\eta_c$ and the associated optimal measurement settings
$\{\hat{A}_{1c},\hat{A}_{2c},\hat{B}_{1c},\hat{B}_{2c}\}$ that achieve
statistical strength $X$. The strategy for finding such solutions
$\{\eta_c, \hat{A}_{1c},\hat{A}_{2c},\hat{B}_{1c},\hat{B}_{2c}\}$ is
as follows: First we start with a set of solutions $\{\eta_{\text{old}},
\hat{A}_{\text{1old}}, \hat{A}_{\text{2old}}, \hat{B}_{\text{1old}}, \hat{B}_{\text{2old}}\}$
having statistical strength $X_{\text{old}}\geq X$. Second we optimize
Eq.~\eqref{statistical strength} over the measurement settings
$\{\hat{A}_1,\hat{A}_2,\hat{B}_1,\hat{B}_2\}$ with fixed detection
efficiency $\eta_{\text{old}}$, which yields new settings
$\{\hat{A}_{\text{1new}}, \hat{A}_{\text{2new}}, \hat{B}_{\text{1new}}, \hat{B}_{\text{2new}}\}$
achieving $S=Y$ ($Y\geq X_{\text{old}}$) for efficiency $\eta_{\text{old}}$. Third,
we decrease the detection efficiency from $\eta_{\text{old}}$ to $\eta_{\text{new}}$
as much as we can without reducing the statistical strength to below $X$,
so that this new set of solutions $\{\eta_{\text{new}}, \hat{A}_{\text{1new}},
\hat{A}_{\text{2new}}, \hat{B}_{\text{1new}}, \hat{B}_{\text{2new}}\}$ has $S=X_{\text{new}}$ with
$X_{\text{new}}$ close to $X$ (within numerical error). We then repeat the
above procedure several times replacing the old with the new solutions,
until we are unable to reduce the efficiency parameter. We thus find
heuristically optimal solutions $\{\eta_c, \hat{A}_{1c}, \hat{A}_{2c},
\hat{B}_{1c}, \hat{B}_{2c}\}$.

\begin{table}[htb!]
\small{
  \caption{Extreme conditions for tests of LR free of the detection
  loophole for photon counters or photon detectors using
  the unbalanced Bell states $|\psi_{\text{uB}}\rangle$ defined in
  Eq.~\eqref{unbalanced Bell state}. The asymptotic behavior when
  $\theta\to 0$ is consistent with results in Ref.~\cite{Vertesi},
  which are shown in the last row. The angle parameters are explained
  in the text.}
  \begin{tabular*}{0.75\textwidth}{@{\extracolsep{\fill}}c c c c c c}
  \hline
  \hline
  $\theta$     &     $\alpha_{\text{1min}}$        &   $\alpha_{\text{2min}}$        &   $\beta_{\text{1min}}$        &    $\beta_{\text{2min}}$         &  $\eta_{\text{min}}$\\
  \hline
  $45^{\circ}$ &     $22.50^{\circ}$    &   $-67.50^{\circ}$   &   $-22.50^{\circ}$  &     $67.50^{\circ}$   &  $82.85\,\%$\\
  $40^{\circ}$ &     $21.28^{\circ}$    &   $-66.89^{\circ}$   &   $-21.28^{\circ}$   &    $66.89^{\circ}$   &  $80.61\,\%$\\
  $35^{\circ}$ &     $19.40^{\circ}$    &   $-65.60^{\circ}$   &   $-19.40^{\circ}$   &    $65.60^{\circ}$   &  $78.50\,\%$\\
  $30^{\circ}$ &     $17.00^{\circ}$    &   $-63.58^{\circ}$   &   $-17.00^{\circ}$   &    $63.58^{\circ}$   &  $76.50\,\%$\\
  $25^{\circ}$ &     $14.21^{\circ}$    &   $-60.72^{\circ}$   &   $-14.21^{\circ}$   &    $60.72^{\circ}$   &  $74.60\,\%$\\
  $20^{\circ}$ &     $11.14^{\circ}$    &   $-56.79^{\circ}$   &   $-11.14^{\circ}$   &    $56.79^{\circ}$   &  $72.81\,\%$\\
  $15^{\circ}$ &     $7.92^{\circ}$    &   $-51.42^{\circ}$   &   $-7.92^{\circ}$   &    $51.42^{\circ}$   &  $71.12\,\%$\\
  $10^{\circ}$ &     $4.70^{\circ}$    &   $-43.88^{\circ}$   &   $-4.70^{\circ}$   &    $43.88^{\circ}$   &  $69.53\,\%$\\
  $5^{\circ}$  &     $1.81^{\circ}$    &   $-32.41^{\circ}$   &   $-1.81^{\circ}$   &    $32.41^{\circ}$   &  $68.06\,\%$\\
  $4^{\circ}$  &     $1.32^{\circ}$    &   $-29.25^{\circ}$   &   $-1.32^{\circ}$   &    $29.25^{\circ}$ &  $67.78\,\%$\\
  $3^{\circ}$  &     $0.87^{\circ}$    &   $-25.55^{\circ}$   &   $-0.87^{\circ}$   &    $25.55^{\circ}$ &  $67.52\,\%$\\
  $2^{\circ}$  &     $0.48^{\circ}$    &   $-21.04^{\circ}$   &   $-0.48^{\circ}$   &    $21.04^{\circ}$ &  $67.27\,\%$\\
  $1^{\circ}$  &     $0.17^{\circ}$    &   $-15.01^{\circ}$   &   $-0.17^{\circ}$   &    $15.01^{\circ}$ &  $67.06\,\%$\\
  $\to 0$  &      $0$    &    $\to -2\theta^{1/2}$   &    $0$   &  $\to 2\theta^{1/2}$     &  $\to 2/3$\\
  \hline
  \hline
  \end{tabular*}
  \label{results for unbalanced Bell states}}
\end{table}

First, we analyze unbalanced Bell states of the form
\begin{equation}
\vert\psi_{\text{uB}}\rangle=\cos(\theta)|H\rangle_A|H\rangle_B+\sin(\theta)|V\rangle_A|V\rangle_B\label{unbalanced Bell state},
\end{equation}
where $\theta \in(0,\pi/4]$. Note that whether there is a relative
phase $e^{i\Delta\phi}$ between the second and first terms of
Eq.~\eqref{unbalanced Bell state} is not important, since Alice can
always adjust her polarization basis, i.e., $|H\rangle_A \to
|H\rangle_A$, and $|V\rangle_A \to e^{-i\Delta\phi}|V\rangle_A$, to
put the state in the above form. In principle, the state
$|\psi_{\text{uB}}\rangle$ can be simulated by postselection on the state
$|\psi_{\text{pB}}\rangle$ [Eq.~\eqref{real state}], although this introduces
a loophole as mentioned earlier. Experimental techniques to prepare
$|\psi_{\text{uB}}\rangle$ without postselection have been demonstrated and
applied to tests of LR~\cite{White,Brida}. Here we calculate the
statistical strength for photon detectors. Photon counters have no
advantage over photon detectors here, because no more than
one photon arrives at Alice's or Bob's detectors. That is, counters
and detectors have the same possibilities for detection outcomes and
with the same probabilities. Our optimization results are summarized
in Table~\ref{results for unbalanced Bell states} and Fig.~\ref{KL
divergence for unbalanced Bell states}. The measurement angle
$\alpha_{i,\,\text{min}}$ (or $\beta_{j,\,\text{min}}$) shown in Table~\ref{results for
unbalanced Bell states} is the angle from the $z$ axis of the
polarization state of an incoming photon that gets reflected at PBS2
(or PBS3) in Fig.~\ref{exp. setup}, where we use the Bloch sphere
representation for this state. By convention, $|H\rangle$ and
$\frac{1}{\sqrt{2}}(|H\rangle+|V\rangle)$ are polarization states
associated with the $z$ and $x$ axes, respectively. In general, we let
the ``unhatted'' form of the measurement setting denote twice the
traceless part of this reflected state's density matrix, or
equivalently, the measurement operator that describes the effect of
the PR, PBS and ideal detector combination on single photon states.
For example, $A_{ic} = \cos(\alpha_{ic})\sigma_z
+\sin(\alpha_{ic})[\cos(\phi_{ic})\sigma_x +
\sin(\phi_{ic})\sigma_y]$. The optimizations show heuristically that
we can take $\phi_{ic}=0$ everywhere; i.e., all the optimal
measurement settings lie in the $(x,z)$ plane of the Bloch sphere, an
observation which has been proven for several special
cases~\cite{Gisin, Popescu2, Scarani}.

From Table~\ref{results for unbalanced Bell states}, we can see that
when the statistical strength $S$ approaches $0$,
$\alpha_{i,\,\text{min}}=-\beta_{i,\,\text{min}}$ for $i=1, 2$. The minimum detection
efficiency $\eta_{\text{min}}$ decreases monotonically with the parameter
$\theta$ in $|\psi_{\text{uB}}\rangle$ and is $82.85\,\%$ when
$\theta=\pi/4$, where the state is a Bell state. It approaches $2/3$
when $\theta$ approaches $0$, where the state is very close to a
product state. These results are consistent with previous
results~\cite{Garg, Eberhard}. From Fig.~\ref{KL divergence for
unbalanced Bell states}, we can see how the optimal statistical
strength increases for $\eta>\eta_{\text{min}}$ and how the input state must
change to achieve this statistical strength. Note that not all
unbalanced Bell states can achieve a given statistical strength level
$S>0$, even for $\eta=1$. For example, for $S\geq 10^{-4}$, the
parameter $\theta$ must be greater than $0.98^{\circ}$. Associated
measurement settings can be found in the tables in the appendix.

\begin{figure}[htb!]
   \includegraphics[scale=0.75, bb=0.5cm 7.3cm 20cm 21cm]{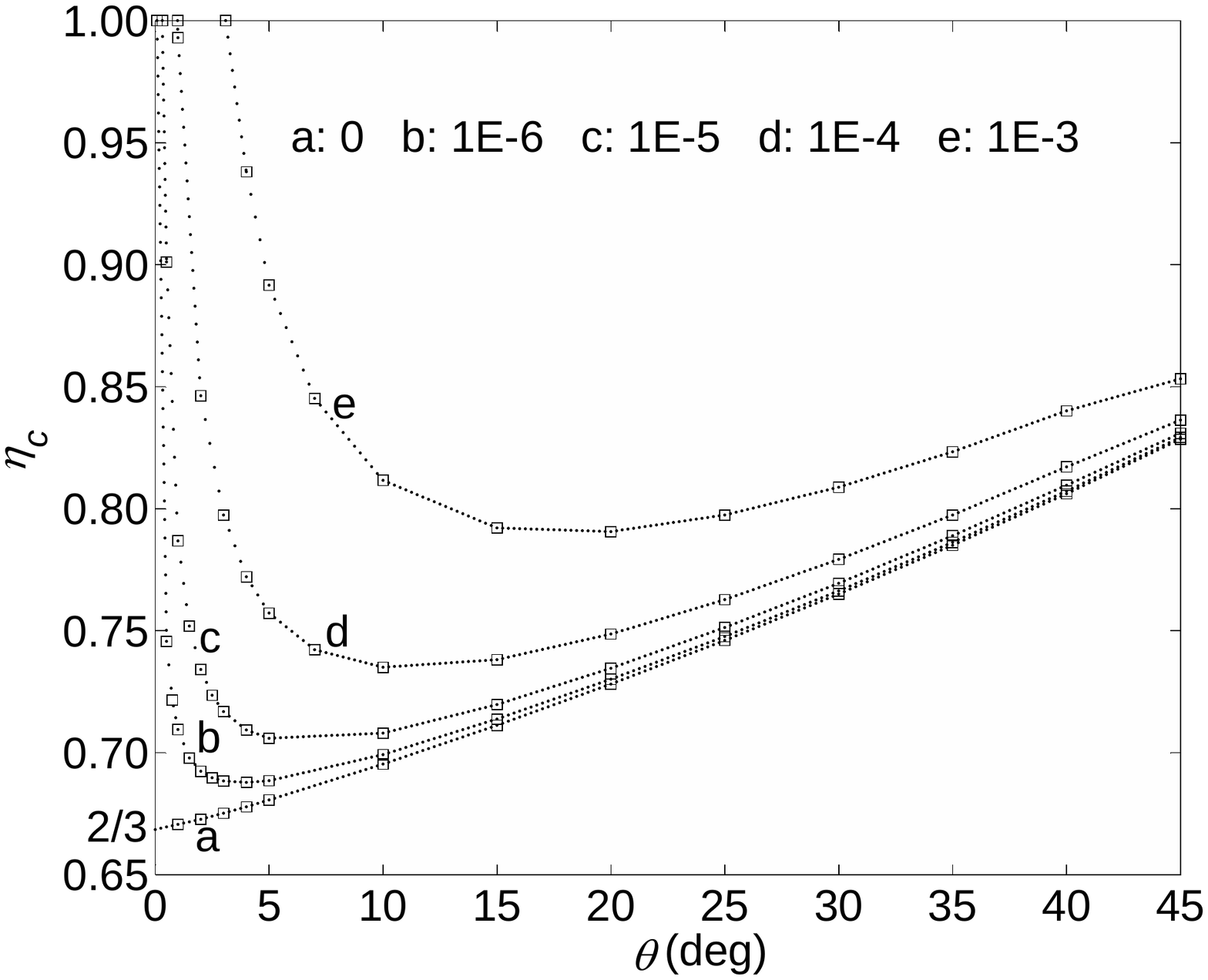}
   \small{
   \caption{Detection efficiency of photon counters or photon
   detectors required for different statistical strength levels $S$ vs
   the parameter $\theta$ [Eq.~\eqref{unbalanced Bell state}]. The
   empty squares show our calculated points, and the dotted lines are linear
   interpolations to guide the eyes. In curve a, the linear
   extrapolation toward $\theta=0$ is shown.}
   \label{KL divergence for unbalanced Bell states}}
\end{figure}

We now consider the pseudo-Bell states of Eq.~\eqref{real state}. Let
$\alpha=\cos(\gamma)$ and $\beta=\sin(\gamma)e^{i\phi}$, then
Eq.~\eqref{real state} can be rewritten as
\begin{align}
|\psi_{\text{pB}}\rangle &=\cos^2(\gamma)|H\rangle_3|H\rangle_4+\sin^2(\gamma)e^{i2\phi}|V\rangle_3|V\rangle_4
         \nonumber \\                                        
                       &\qquad +\cos(\gamma)\sin(\gamma)e^{i\phi}(|H\rangle_3|V\rangle_3+|H\rangle_4|V\rangle_4),\label{real state2}
\end{align}
where $\gamma \in(0, \pi/4]$, and $\phi \in[0, 2\pi)$. We can prepare
different pseudo-Bell states by changing the values of both $\gamma$
and $\phi$. However, for a given $\gamma$, as the following discussion
shows, the optimal statistical strength $S$ is the same regardless of
the value of $\phi$. In the test of LR as shown in Fig.~\ref{exp.
setup}, Alice's and Bob's measurements are restricted to polarization
rotation followed by photon counting. They cannot detect coherences
between any two of the first two, the third, and the last terms in the
state $|\psi_{\text{pB}}\rangle$ as written in Eq.~\eqref{real state2},
because these terms correspond to different photon-number-distribution
subspaces. Hence, the measurement outcomes determined by
$|\psi_{\text{pB}}\rangle$ are equivalent to the outcomes given by a mixture
of the following two states:
\begin{equation}
|\psi_1\rangle\langle\psi_1|,\; \textrm{with}\; |\psi_1\rangle \propto \cos^2(\gamma)|H\rangle_3|H\rangle_4+\sin^2(\gamma)e^{i2\phi}|V\rangle_3|V\rangle_4,
\end{equation}
and
\begin{equation}
\rho_2 \propto |H\rangle_3 |V\rangle_3 \hspace{0.2cm} _3\langle H|_3\langle V|+|H\rangle_4|V\rangle_4 \hspace{0.2cm} _4\langle H| _4\langle V|.
\end{equation}
Since the state $|\psi_1\rangle$ can be written in the form
$\vert\psi_{\text{uB}}\rangle$ as in Eq.~\eqref{unbalanced Bell state} by
changing the mode labels and the state bases, the measurement outcomes
attributable to $|\psi_1\rangle$ can reveal a violation of LR
when $\gamma \in(0, \pi/4]$, as our earlier results show. But
$\rho_2$ is a separable state and so the outcomes attributable to
$\rho_2$ can be explained by LR no matter what the
measurement settings $\{\hat A_1, \hat A_2, \hat B_1, \hat B_2\}$ are.
Hence, in a test of LR, the information about whether LR is
or is not violated is conveyed only by the outcomes from
$|\psi_1\rangle$, while the state $\rho_2$ acts as noise. Based on
these considerations and the earlier arguments about being able to
eliminate a potential phase in $\vert\psi_{\text{uB}}\rangle$, we do not need
to consider different phases $\phi$ in the pseudo-Bell state
$|\psi_{\text{pB}}\rangle$ when calculating the optimal statistical strength
$S$, so we can choose a fixed value, such as $\phi=0$. Moreover, we
determined heuristically by extended optimizations in selected cases
that the optimal measurement settings $\{\hat A_{1c}, \hat A_{2c},
\hat B_{1c}, \hat B_{2c}\}$ can be chosen to lie in the $(x, z)$ plane
of the Bloch sphere, just like for $\vert\psi_{\text{uB}}\rangle$. Taking
these observations into account reduces the number of free parameters
and speeds up the general calculations.

The optimization results for pseudo-Bell states are summarized in
Table~\ref{results for real states} and Fig.~\ref{KL divergence for
real states with photon counters and photon detectors}. Similar to
unbalanced Bell states, Table~\ref{results for real states} shows that
when the statistical strength $S$ approaches $0$,
$\alpha_{i,\,\text{min}}=-\beta_{i,\,\text{min}}$ for $i=1, 2$. Figure~\ref{KL divergence for
real states with photon counters and photon detectors} shows that there is 
a lower bound on the state parameter $\gamma$ to achieve a nonzero 
statistical strength level $S$. Measurement settings for
the data shown in Fig.~\ref{KL divergence for real states with photon
counters and photon detectors} are given in the appendix.

\begin{table}[htb!]
\small{
  \caption{Extreme conditions for tests of LR free of the detection
  loophole for photon counters and photon detectors using the
  pseudo-Bell states of Eq.~\eqref{real state2}. The angle parameters are explained
  in the text.}
  \begin{tabular*}{1.00\textwidth}{ @{\extracolsep{\fill}} c  c c c c c  c c c c c}
  \hline
  \hline
              &  \multicolumn{5}{c}{Photon counter} & \multicolumn{5}{c}{Photon detector}\\
  \hline
  $\gamma$    &  $\alpha_{\text{1min}}$      &   $\alpha_{\text{2min}}$       &   $\beta_{\text{1min}}$        &  $\beta_{\text{2min}}$        &    $\eta_{\text{min}}$   & $\alpha_{\text{1min}}$   & $\alpha_{\text{2min}}$      & $\beta_{\text{1min}}$       &  $\beta_{\text{2min}}$    & $\eta_{\text{min}}$\\
  \hline
  
  $45^{\circ}$  &  $22.50^{\circ}$  &   $-67.50^{\circ}$  &   $-22.50^{\circ}$  &  $67.50^{\circ}$   &    $90.62\,\%$  & $11.64^{\circ}$ & $-63.88^{\circ}$  & $-11.64^{\circ}$ &  $63.88^{\circ}$  &   $92.23\,\%$\\
  
  $40^{\circ}$  &  $20.49^{\circ}$  &   $-66.01^{\circ}$  &   $-20.49^{\circ}$  &  $66.01^{\circ}$   &    $89.71\,\%$  & $11.08^{\circ}$ & $-62.79^{\circ}$  & $-11.08^{\circ}$ &  $62.79^{\circ}$  &   $91.31\,\%$\\
  
  $35^{\circ}$  &  $16.76^{\circ}$  &   $-62.14^{\circ}$  &   $-16.76^{\circ}$  &  $62.14^{\circ}$   &    $89.78\,\%$  & $9.79^{\circ}$ & $-59.60^{\circ}$  & $-9.79^{\circ}$ &  $59.60^{\circ}$  &   $91.11\,\%$\\
  
  $30^{\circ}$  &  $12.32^{\circ}$  &   $-56.16^{\circ}$  &   $-12.32^{\circ}$  &  $56.16^{\circ}$   &    $90.80\,\%$  & $7.93^{\circ}$ & $-54.42^{\circ}$  & $-7.93^{\circ}$ &  $54.42^{\circ}$  &   $91.71\,\%$\\
  
  $25^{\circ}$  &  $8.00^{\circ}$  &   $-48.43^{\circ}$  &   $-8.00^{\circ}$  &  $48.43^{\circ}$   &    $92.57\,\%$  & $5.73^{\circ}$ & $-47.46^{\circ}$  & $-5.73^{\circ}$ &  $47.46^{\circ}$  &   $93.05\,\%$\\
  
  $20^{\circ}$  &  $4.43^{\circ}$  &   $-39.49^{\circ}$  &   $-4.43^{\circ}$  &  $39.49^{\circ}$   &    $94.71\,\%$  & $3.53^{\circ}$ & $-39.09^{\circ}$  & $-3.53^{\circ}$ &  $39.09^{\circ}$  &   $94.89\,\%$\\
  
  $15^{\circ}$  &  $1.96^{\circ}$  &   $-29.88^{\circ}$  &   $-1.96^{\circ}$  &  $29.88^{\circ}$   &    $96.81\,\%$  & $1.68^{\circ}$ & $-29.76^{\circ}$  & $-1.68^{\circ}$ &  $29.76^{\circ}$  &   $96.85\,\%$\\
  
  $10^{\circ}$  &  $0.59^{\circ}$  &   $-19.98^{\circ}$    &   $-0.59^{\circ}$  &  $19.98^{\circ}$     &    $98.52\,\%$  & $0.54^{\circ}$ & $-19.96^{\circ}$  & $-0.54^{\circ}$ &  $19.96^{\circ}$    &   $98.53\,\%$\\
  
  $5^{\circ}$   &  $0.07^{\circ}$  &   $-10.00^{\circ}$    &   $-0.07^{\circ}$  &  $10.00^{\circ}$     &    $99.63\,\%$  & $0.07^{\circ}$ & $-10.00^{\circ}$  & $-0.07^{\circ}$ &  $10.00^{\circ}$    &   $99.63\,\%$ \\
  
  \hline
  \hline
  \end{tabular*}
  \label{results for real states}}
\end{table}

\begin{figure}[htb!]
   \includegraphics[scale=0.75, bb=0.5cm 7.7cm 20cm 21cm]{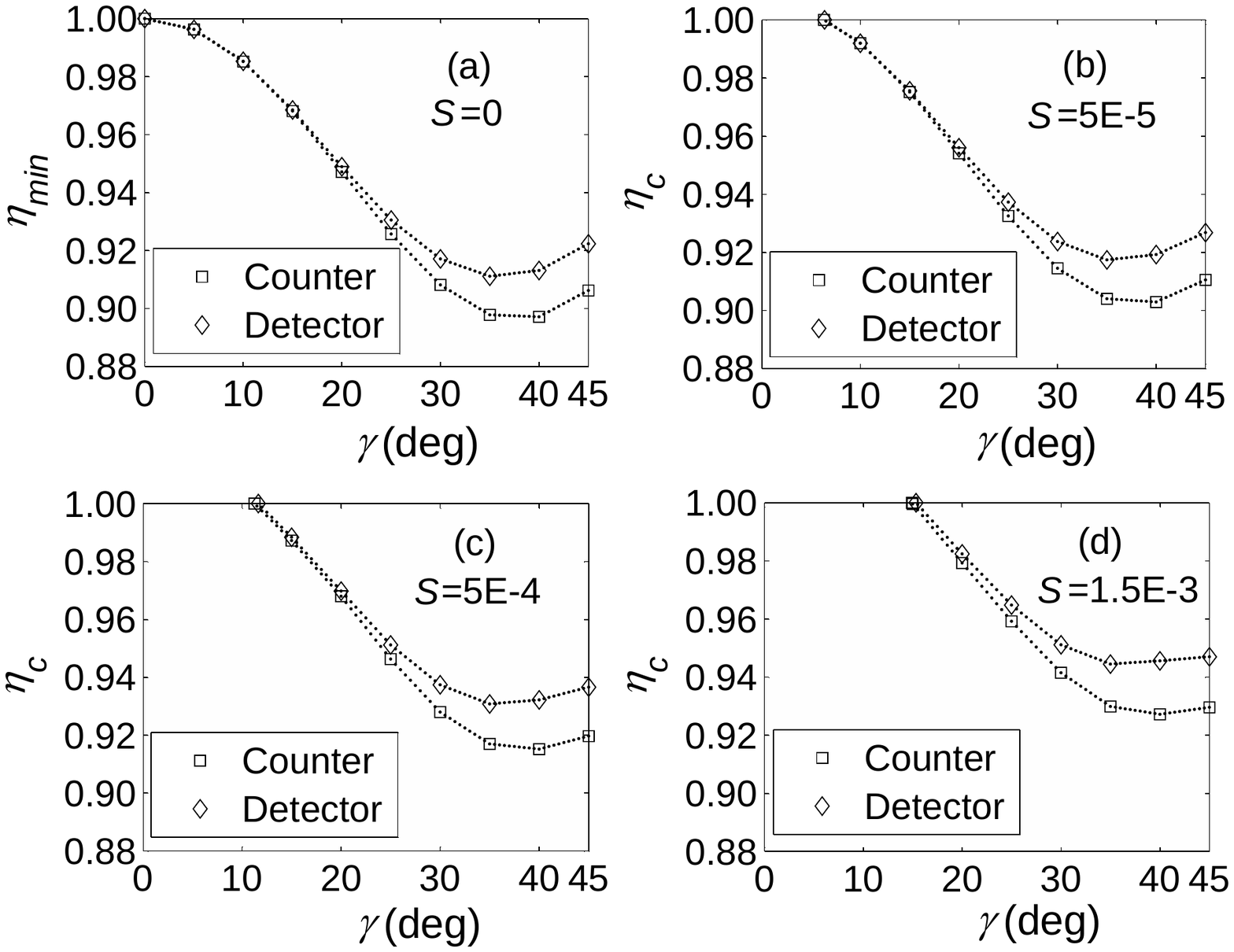}
   \small{
   \caption{Detection efficiencies of photon counters and photon
   detectors required for different statistical strength levels $S$ vs
   the parameter $\gamma$ of the pseudo-Bell state of Eq.~\eqref{real
   state2}: (a) $S=0$, (b) $S=$ 5E-5, (c) $S=$ 5E-4, and (d) $S=$
   1.5E-3. The calculated points are labeled by squares for photon
   counters and by diamonds for photon detectors, and the dotted lines 
   are linear interpolations to guide the eyes. }
   \label{KL divergence for real states with photon counters and photon detectors}}
\end{figure}

Table~\ref{results for real states} and Fig.~\ref{KL divergence for
real states with photon counters and photon detectors} (a) show that
the minimum detection efficiency $\eta_{\text{min}}$ required to close the
detection loophole achieves its minimum in the interior of the domain,
in contrast to what was found for the case of unbalanced Bell states.
We might have expected this behavior based on the following
observations: First, with respect to the detector setups used, the
state $\vert\psi_{\text{pB}}\rangle$ can be thought of as the state
$|\psi_{\text{uB}}\rangle$ with noise, as pointed out above, and second, the
violation of LR given by $|\psi_{\text{uB}}\rangle$ is very sensitive to
noise, particularly when $\theta$ [Eq.~\eqref{unbalanced Bell state}]
is small~\cite{Brunner}. Figure~\ref{KL divergence for real states
with photon counters and photon detectors} (a) also suggests that any
pseudo-Bell state $\vert\psi_{\text{pB}}\rangle$ can violate LR using
counters or detectors with sufficient efficiency.

When we look at the minimum detection efficiency required to achieve a
given statistical strength level $S$, the efficiencies of photon
counters and photon detectors are notably different, showing the
utility of the additional information available with photon counters.
The advantage of photon counters is most notable for $\gamma$ between
approximately $35^{\circ}$ and $45^{\circ}$. In particular, the minimum
detection efficiency $\eta_{\text{min}}$ is $89.71\,\%$ for photon
counters and $91.11\,\%$ for photon detectors and is achieved for
$\gamma$ in this range. Loosely speaking, this advantage is because
photon counters are better at differentiating between measurement
outcomes contributed by the entangled ($|\psi_1\rangle$) and
unentangled ($\rho_2$) parts of the state $\vert\psi_{\text{pB}}\rangle$.

A comparison of Figs.~\ref{KL divergence for unbalanced Bell states}
and \ref{KL divergence for real states with photon counters and
photon detectors} suggests that higher efficiencies are required to
achieve given statistical strengths with $|\psi_{\text{pB}}\rangle$ than with
$|\psi_{\text{uB}}\rangle$. This again can be attributed to the noise added
by $\rho_2$ to measurement outcomes, which reduces the statistical
strength considerably. As an explicit example, consider the optimal
statistical strengths $S^{(1)}$ or $S^{(2)}$ achievable with
\begin{equation}
|\psi_{\text{uB}}(\theta=\pi/4)\rangle=\frac{1}{\sqrt{2}}(|H\rangle_A|H\rangle_B+|V\rangle_A|V\rangle_B),
\end{equation}
or with
\begin{equation}
|\psi_{\text{pB}}(\gamma=\pi/4,\phi=0)\rangle=\frac{1}{2}(|H\rangle_3|H\rangle_4+|V\rangle_3|V\rangle_4+|H\rangle_3|V\rangle_3+|H\rangle_4|V\rangle_4).
\end{equation}
We find that $S^{(1)}=2S^{(2)}\approx 0.04627$ for perfect photon
counters. The ratio can be explained by observing that half of the
measurement outcomes of $|\psi_{\text{pB}}(\gamma=\pi/4,\phi=0)\rangle$ are
from the separable $\rho_2$.

\section{Conclusion}
\label{sect:conclusion}

We have demonstrated a method to measure the statistical strength of
tests of LR that is based on the KL divergence from the predicted
experimental frequencies to the best prediction given by LR. This
method helps to design a loophole-free test of LR and quantifies the
confidence in violation of LR for sufficiently large experimental data
sets. We used the method to determine optimal statistical strengths of
tests of LR using a typical detector setup for polarized photon pairs with inefficient
detectors. We considered both ideal unbalanced Bell states and
pseudo-Bell states obtained by combining independent polarized photons
on a polarizing beam splitter. Creating the latter can be
easier~\cite{Shih,Ou,Kiess}, but observing a violation of LR requires
higher detection efficiencies. Our calculations show that with
pseudo-Bell states, we can close the detection loophole with a minimum
detection efficiency of $89.71\,\%$ using photon counters, or
$91.11\,\%$ using photon detectors. For unbalanced Bell states, we
confirmed previous calculations~\cite{Eberhard} showing that
violations of LR are possible at detection efficiencies above $2/3$.
Furthermore, we numerically exhibited the relationships between state
parameters and minimum detection efficiencies needed to achieve given
levels of statistical strength. Given that the current roadblock for
performing loophole-free tests of LR with photons is detection
inefficiency rather than the difficulty of obtaining an entangled
source, we cannot recommend using the pseudo-Bell state for such an
experiment. 

In current experiments based on spontaneous parametric down-conversion
to produce entangled photon pairs, we must consider other sources of
potentially unwanted measurement outcomes. Such sources include dark
counts and the generation of more than one photon
pair~\cite{Wasilewski,Lvovsky}. The latter effect can be quite
noticeable, particularly for the brighter, more strongly pumped
sources. Further work is required to analyze the consequences of these
effects for statistical strength. It is also desirable to obtain
rigorous confidence levels for the rejection of LR with moderately
sized data sets. Such levels could improve on measures derived from
experimental standard deviations of Bell-inequality violation.

\begin{acknowledgments}
We thank Kevin Coakley for providing information about maximum
likelihood algorithms, and Adam Meier, Bryan Eastin and Mike Mullan
for discussions and comments. This paper is a contribution of the
National Institute of Standards and Technology and not subject to U.S.
copyright.
\end{acknowledgments}

\appendix*
\section{Optimization results}

Using code written in Octave~\footnote{The use of trade names is for
informational purposes only and does not imply endorsement by NIST.},
which is available by request, we find the results as shown in the
following tables. In these tables, the units of the columns labeled
\verb+theta+, \verb+gamma+, \verb+A_1+, \verb+A_2+, \verb+B_1+, and
\verb+B_2+ are degrees ($\circ$). The column labeled \verb+theta+ (or
\verb+gamma+) contains the values of the parameter $\theta$ in the
state
$\vert\psi_{\text{uB}}\rangle=\cos(\theta)|H\rangle_A|H\rangle_B+\sin(\theta)|V\rangle_A|V\rangle_B$
(or the value of the parameter $\gamma$ in the state
$|\psi_{\text{pB}}\rangle=\cos^2(\gamma)|H\rangle_3|H\rangle_4+\sin^2(\gamma)|V\rangle_3|V\rangle_4+
\cos(\gamma)\sin(\gamma)(|H\rangle_3|V\rangle_3+|H\rangle_4|V\rangle_4)$).
The columns labeled \verb+A_1+ and \verb+A_2+ contain the two optimal
measurement setting angles for Alice, while the columns labeled
\verb+B_1+ and \verb+B_2+ contain the two optimal measurement setting
angles for Bob. The columns labeled \verb+eta_1+ and \verb+eta_2+ give
the detection efficiencies required to achieve the statistical
strengths in the columns labeled \verb+S_1+ and \verb+S_2+,
respectively. Due to limits of numerical accuracy, we cannot find the
exact detection efficiency $\eta_c$ required to achieve a specified
statistical strength level $S$. Up to the $10^{-4}$ level, we list the
best two detection efficiencies, which are closest to $\eta_c$. Using
\verb+eta_1+, the statistical strength of a test of LR is a little
higher than the specified level, while using \verb+eta_2+, the
statistical strength is a little lower than the specified level. In
the plots of Fig.~\ref{KL divergence for unbalanced Bell states} and
Fig.~\ref{KL divergence for real states with photon counters and
photon detectors}, we use \verb+eta_1+ or \verb+eta_2+, according to
which of them gives a statistical strength closer to the specified
level. Also, for calculations of the minimum detection efficiency
at $0+\epsilon$ statistical strength, we truncate the statistical
strength to $0$ when it is numerically calculated to be less than
$10^{-9}$ or $10^{-10}$, depending on the situation.

\subsection{Results for states $\vert\psi_{\text{uB}}\rangle=\cos(\theta)|H\rangle_A|H\rangle_B+\sin(\theta)|V\rangle_A|V\rangle_B$ using photon counters or photon detectors}


\small
\begin{verbatim}
Statistical strength~=0								
theta	 A_1	 A_2	  B_1     B_2    eta_1      S_1       eta_2      S_2
 45	22.50	-67.50	-22.50	 67.50	 82.85%   8.66E-009   82.84%   1.44E-013  
 40	21.28	-66.89	-21.28	 66.89	 80.61%	  1.47E-010   80.60%   3.18E-015
 35	19.40	-65.60	-19.40	 65.60	 78.50%	  3.97E-009   78.49%   1.33E-013
 30	17.00	-63.58	-17.00	 63.58	 76.50%	  5.48E-009   76.49%   3.62E-011
 25	14.21	-60.72	-14.21	 60.72	 74.60%	  1.48E-009   74.59%   1.66E-013
 20	11.14	-56.79	-11.14	 56.79	 72.81%	  2.91E-009   72.80%   5.56E-011
 15	7.92	-51.42	-7.92	 51.42	 71.12%	  3.84E-009   71.11%   6.85E-010
 10	4.70	-43.88	-4.70	 43.88	 69.53%	  2.56E-009   69.52%   7.10E-010
  5	1.81	-32.41	-1.81	 32.41	 68.06%	  1.67E-009   68.05%   8.37E-010
  4	1.32	-29.25	-1.32	 29.25	 67.78%	  1.21E-009   67.77%   6.40E-010
  3	0.87	-25.55	-0.87	 25.55	 67.52%	  1.48E-009   67.51%   9.78E-010
  2	0.48	-21.04	-0.48	 21.04	 67.27%	  1.31E-009   67.26%   9.85E-010
  1	0.17	-15.01	-0.17	 15.01	 67.06%	  1.00E-009   67.05%   8.57E-010
 								
								
Statistical strength~=1E-6								
theta	 A_1	 A_2	  B_1     B_2    eta_1      S_1       eta_2      S_2
 45	22.50	-67.50	-22.50	 67.50	 82.93%   1.24E-006   82.92%   9.71E-007
 40	21.28	-66.89	-21.28	 66.89	 80.72%	  1.00E-006   80.71%   8.30E-007
 35	19.41	-65.60	-19.41	 65.60	 78.62%	  1.03E-006   78.61%   8.79E-007
 30	17.02	-63.59	-17.02	 63.59	 76.64%   1.09E-006   76.63%   9.48E-007
 25	14.23	-60.73	-14.23	 60.73	 74.77%   1.07E-006   74.76%   9.53E-007   
 20	11.15	-56.80	-11.15	 56.80	 73.01%	  1.01E-006   73.00%   9.18E-007
 15	7.93	-51.43	-7.93	 51.43	 71.39%   1.07E-006   71.38%   1.00E-006   
 10	4.72	-43.89	-4.72	 43.89	 69.93%   1.04E-006   69.92%   9.88E-007   
  5	1.82	-32.42	-1.82	 32.42	 68.86%   1.02E-006   68.85%   9.96E-007   
  4	1.33	-29.25	-1.33	 29.25	 68.78%	  1.01E-006   68.77%   9.90E-007
  3	0.88	-25.55	-0.88	 25.55	 68.84%	  1.00E-006   68.83%   9.86E-007
2.5     0.68    -23.43  -0.68    23.43   68.98%   1.01E-006   68.97%   9.94E-007
  2	0.49	-21.05	-0.49	 21.05	 69.25%	  1.01E-006   69.24%   9.96E-007
1.5     0.32    -18.31  -0.32    18.31   69.78%   1.00E-006   69.77%   9.94E-007	
  1	0.18	-15.01	-0.18	 15.01	 70.96%	  1.00E-006   70.95%   9.98E-007
0.75    0.12    -13.03  -0.12    13.03   72.17%   1.00E-006   72.16%   9.98E-007    
0.5     0.07    -10.66  -0.07    10.66   74.56%   1.00E-006   74.55%   9.98E-007
0.08	0.00	 90.00  -0.16	 0.16	100.00%	  1.00E-006
								
								
Statistical strength~=1E-5								
theta	 A_1	 A_2	  B_1     B_2    eta_1      S_1       eta_2      S_2
 45	22.50	-67.50	-22.50	 67.50	 83.10%   1.08E-005   83.09%   9.95E-006   
 40	21.30	-66.90	-21.30	 66.90	 80.96%	  1.00E-005   80.95%   9.46E-006
 35	19.43	-65.62	-19.43	 65.62	 78.89%	  1.01E-005   78.88%   9.56E-006
 30	17.05	-63.61	-17.05	 63.61	 76.95%   1.02E-005   76.94%   9.79E-006   
 25	14.26	-60.76	-14.26	 60.76	 75.14%   1.03E-005   75.13%   9.96E-006   
 20	11.20	-56.84	-11.20	 56.84	 73.47%   1.03E-005   73.46%   9.99E-006   
 15	7.97	-51.47	-7.97	 51.47	 71.98%   1.01E-005   71.97%   9.90E-006   
 10	4.75	-43.92	-4.75	 43.92	 70.81%   1.01E-005   70.80%   9.92E-006   
  5	1.85	-32.45	-1.85	 32.45	 70.60%   1.01E-005   70.59%   9.99E-006   
  4	1.35	-29.27	-1.35	 29.27	 70.94%   1.00E-005   70.93%   9.97E-006
  3	0.90	-25.57	-0.90	 25.57	 71.69%	  1.00E-005   71.68%   9.98E-006
2.5     0.69    -23.45  -0.69    23.45   72.36%   1.00E-005   72.35%   9.98E-006
  2	0.50	-21.06	-0.50	 21.06	 73.41%	  1.00E-005   73.40%   9.98E-006
1.5     0.33    -18.32  -0.33    18.32   75.20%   1.00E-005   75.19%   1.00E-005
  1	0.19	-15.02	-0.19	 15.02	 78.69%	  1.00E-005   78.68%   9.99E-006
0.5     0.00     90.00  -0.95    0.95    90.11%   1.00E-005   90.10%   9.99E-006
0.31	0.00	 90.00 	-0.61	 0.61	100.00%	  1.00E-005 
								
								
Statistical strength~=1E-4								
theta	 A_1	 A_2	  B_1     B_2    eta_1      S_1       eta_2      S_2
 45	22.50	-67.50	-22.50	 67.50	 83.63%	  1.01E-004   83.62%   9.83E-005
 40	21.34	-66.94	-21.34	 66.94	 81.71%	  1.00E-004   81.70%   9.82E-005
 35	19.51	-65.69	-19.51	 65.69	 79.74%	  1.01E-004   79.73%   9.90E-005
 30	17.16	-63.71	-17.16	 63.71	 77.92%	  1.00E-004   77.91%   9.91E-005
 25	14.39	-60.87	-14.39	 60.87	 76.28%   1.01E-004   76.27%   9.95E-005   
 20	11.33	-56.96	-11.33	 56.96	 74.87%   1.01E-004   74.86%   9.98E-005   
 15	8.09	-51.58	-8.09	 51.58	 73.81%	  1.00E-004   73.80%   9.94E-005
 10	4.86	-44.03	-4.86	 44.03	 73.50%	  1.00E-004   73.49%   9.96E-005
  7     3.03    -37.82  -3.03    37.82   74.22%   1.00E-004   74.21%   9.99E-005
  5	1.92	-32.52	-1.92	 32.52	 75.73%   1.00E-004   75.72%   9.99E-005   
  4	1.41	-29.34	-1.41	 29.34	 77.21%	  1.00E-004   77.20%   9.99E-005
  3	0.95	-25.64	-0.95	 25.64	 79.74%	  1.00E-004   79.73%   9.98E-005
  2	0.54	-21.12	-0.54	 21.12	 84.64%	  1.00E-004   84.63%   1.00E-004
  1	0.00	 90.00  -1.98    1.98	 99.30%	  1.00E-004   99.29%   1.00E-004 
0.98	0.00	 90.00  -1.93    1.93	100.00%	  1.00E-004
								
								
Statistical strength~=1E-3								
theta	 A_1	 A_2	  B_1     B_2    eta_1      S_1       eta_2      S_2
 45	22.50	-67.50	-22.50	 67.50	 85.33%	  1.01E-003   85.32%   9.99E-004
 40	21.47	-67.06	-21.47	 67.06	 84.02%   1.01E-003   84.01%   1.00E-003   
 35	19.77	-65.90	-19.77	 65.90	 82.33%	  1.00E-003   82.32%   9.95E-004
 30	17.50	-63.99	-17.50	 63.99	 80.88%	  1.00E-003   80.87%   9.97E-004
 25	14.79	-61.22	-14.79	 61.22	 79.74%	  1.00E-003   79.73%   9.98E-004
 20	11.74	-57.34	-11.74	 57.34	 79.06%	  1.00E-003   79.05%   9.97E-004
 15	8.48	-51.97	-8.48	 51.97	 79.21%	  1.00E-003   79.20%   9.99E-004
 10	5.18	-44.37	-5.18	 44.37	 81.17%   1.00E-003   81.16%   9.99E-004
  7     3.28    -38.13  -3.28    38.13   84.52%   1.00E-003   84.51%   9.99E-004   
  5	2.15	-32.89	-2.15	 32.89	 89.16%	  1.00E-003   89.15%   1.00E-003
  4	1.68	-29.59	-1.68	 29.59	 93.81%	  1.00E-003   93.80%   1.00E-003
3.09	0.00	 90.00	-6.10	 6.10	100.00%	  1.00E-003

\end{verbatim}

\subsection{Results for states $|\psi_{\text{pB}}\rangle=\cos^2(\gamma)|H\rangle_3|H\rangle_4+\sin^2(\gamma)|V\rangle_3|V\rangle_4+ \cos(\gamma)\sin(\gamma)(|H\rangle_3|V\rangle_3+|H\rangle_4|V\rangle_4)$ using photon counters}
                       
\small
\begin{verbatim}
Statistical strength~=0								
gamma	 A_1	 A_2	  B_1     B_2    eta_1      S_1       eta_2      S_2
 45	22.50	-67.50	-22.50	 67.50	 90.62%	  3.61E-009   90.61%   2.02E-014
 40	20.49	-66.01	-20.49	 66.01	 89.71%	  8.90E-009   89.70%   8.94E-014
 35	16.76	-62.14	-16.76	 62.14	 89.78%	  4.24E-009   89.77%   3.63E-014
 30	12.32	-56.16	-12.32	 56.16	 90.80%	  2.55E-010   90.79%   1.18E-014
 25	8.00	-48.43	-8.00	 48.43	 92.57%	  8.08E-009   92.56%   6.21E-013
 20	4.43	-39.49	-4.43	 39.49	 94.71%	  8.13E-009   94.70%   1.08E-012
 15	1.96	-29.88	-1.96	 29.88	 96.81%	  4.08E-009   96.80%   5.21E-014
 10	0.59	-19.98	-0.59	 19.98	 98.52%	  1.36E-010   98.51%   5.54E-015
  5	0.07	-10.00	-0.07	 10.00	 99.63%	  5.76E-009   99.62%   1.77E-013
 
								
Statistical strength~=5E-5								
gamma	 A_1	 A_2	  B_1     B_2    eta_1      S_1       eta_2      S_2
 45	22.50	-67.50	-22.50	 67.50	 91.05%	  5.11E-005   91.04%   4.88E-005
 40	20.53	-66.05	-20.53	 66.05	 90.30%   5.11E-005   90.29%   4.94E-005   
 35	16.81	-62.20	-16.81	 62.20	 90.40%	  5.06E-005   90.39%   4.89E-005
 30	12.34	-56.21	-12.34	 56.21	 91.46%   5.12E-005   91.45%   4.97E-005   
 25	7.98	-48.45	-7.98	 48.45	 93.25%	  5.06E-005   93.24%   4.91E-005
 20	4.39	-39.49	-4.39	 39.49	 95.42%   5.15E-005   95.41%   5.00E-005   
 15	1.91	-29.87	-1.91	 29.87	 97.52%	  5.02E-005   97.51%   4.87E-005
 10	0.56	-19.97	-0.56	 19.97	 99.20%	  5.01E-005   99.19%   4.84E-005
6.30	0.00	 90.00	-1.38	 1.38	100.00%	  5.00E-005
 
								
Statistical strength~=5E-4								
gamma	 A_1	 A_2	  B_1     B_2    eta_1      S_1       eta_2      S_2
 45	22.50	-67.50	-22.50	 67.50	 91.98%   5.06E-004   91.97%   4.99E-004   
 40	0.00	 90.00	-42.54	 42.54	 91.53%   5.03E-004   91.52%   4.97E-004   
 35	16.96	-62.38	-16.96	 62.38	 91.70%	  5.02E-004   91.69%   4.97E-004
 30	12.42	-56.35	-12.42	 56.35	 92.81%   5.05E-004   92.80%   5.00E-004   
 25	7.96	-48.53	-7.96	 48.53	 94.64%   5.04E-004   94.63%   4.99E-004   
 20	4.30	-39.51	-4.30	 39.51	 96.80%	  5.03E-004   96.79%   4.97E-004
 15	0.00	 90.00	-7.99	 7.99	 98.73%   5.03E-004   98.72%   4.97E-004
11.26	0.00	 90.00	-4.49	 4.49	100.00%	  5.00E-004
  						
													
Statistical strength~=1.5E-3								
gamma	 A_1	 A_2	  B_1     B_2    eta_1      S_1       eta_2      S_2
 45	22.50	-67.50	-22.50	 67.50	 92.97%   1.51E-003   92.96%   1.49E-003   
 40	0.00	 90.00	-42.81	 42.81	 92.73%   1.51E-003   92.72%   1.50E-003   
 35	0.00	 90.00	-37.46	 37.46	 93.00%   1.51E-003   92.99%   1.50E-003   
 30	12.54	-56.55	-12.54	 56.55	 94.16%   1.51E-003   94.15%   1.50E-003   
 25	0.00	 90.00	-21.93	 21.93	 95.92%	  1.50E-003   95.91%   1.49E-003
 20	0.00	 90.00	-14.44	 14.44	 97.94%   1.51E-003   97.93%   1.50E-003   
 15	0.00	 90.00	-8.11	 8.11	 99.97%   1.51E-003   99.96%   1.50E-003
14.92	0.00	 90.00	-8.01	 8.01	100.00%	  1.50E-003
   
\end{verbatim}

\subsection{Results for states $|\psi_{\text{pB}}\rangle=\cos^2(\gamma)|H\rangle_3|H\rangle_4+\sin^2(\gamma)|V\rangle_3|V\rangle_4+ \cos(\gamma)\sin(\gamma)(|H\rangle_3|V\rangle_3+|H\rangle_4|V\rangle_4)$ using photon detectors}

\small
\begin{verbatim}
Statistical strength~=0								
gamma	 A_1	 A_2	  B_1     B_2    eta_1      S_1       eta_2      S_2
 45	11.64	-63.88	-11.64	 63.88	 92.23%	  1.96E-009   92.22%   2.57E-014
 40	11.08	-62.79	-11.08	 62.79	 91.31%	  4.10E-009   91.30%   7.36E-014
 35	9.79	-59.60	-9.79	 59.60	 91.11%	  7.81E-009   91.10%   3.78E-013
 30	7.93	-54.42	-7.93	 54.42	 91.71%	  3.60E-009   91.70%   7.68E-014
 25	5.73	-47.46	-5.73	 47.46	 93.05%	  1.20E-009   93.04%   2.56E-014
 20	3.53	-39.09	-3.53	 39.09	 94.89%	  3.16E-010   94.88%   1.03E-014
 15	1.68	-29.76	-1.68	 29.76	 96.85%	  2.71E-010   96.84%   7.47E-015
 10	0.54	-19.96	-0.54	 19.96	 98.53%	  3.26E-009   98.52%   3.30E-014
  5	0.07	-10.00	-0.07	 10.00	 99.63%	  5.66E-009   99.62%   1.44E-013


				
Statistical strength~=5E-5								
gamma	 A_1	 A_2	  B_1     B_2    eta_1      S_1       eta_2      S_2
 45	11.62	-63.89	-11.62	 63.89	 92.69%   5.13E-005   92.68%   4.91E-005   
 40	11.06	-62.85	-11.06	 62.85	 91.93%   5.11E-005   91.92%   4.94E-005   
 35	9.78	-59.69	-9.78	 59.69	 91.75%   5.12E-005   91.74%   4.96E-005   
 30	7.92	-54.50	-7.92	 54.50	 92.37%	  5.05E-005   92.36%   4.90E-005
 25	5.70	-47.51	-5.70	 47.51	 93.74%   5.11E-005   93.73%   4.96E-005   
 20	3.49	-39.10	-3.49	 39.10	 95.60%	  5.07E-005   95.59%   4.92E-005
 15	1.67	-29.76	-1.67	 29.76	 97.57%   5.12E-005   97.56%   4.97E-005   
 10	0.52	-19.96	-0.52	 19.96	 99.21%   5.11E-005   99.20%   4.94E-005
6.36	0.00	 90.00	-1.40	 1.40	100.00%	  5.00E-005
   
								
								
Statistical strength~=5E-4								
gamma	 A_1	 A_2	  B_1     B_2    eta_1      S_1       eta_2      S_2
 45	11.58	-63.92	-11.58	 63.92	 93.67%   5.04E-004   93.66%   4.97E-004   
 40	11.00	-63.04	-11.00	 63.04	 93.23%   5.02E-004   93.22%   4.97E-004   
 35	9.74	-59.93	-9.74	 59.93	 93.09%   5.04E-004   93.08%   4.99E-004   
 30	7.88	-54.71	-7.88	 54.71	 93.74%	  5.02E-004   93.73%   4.97E-004
 25	5.64	-47.66	-5.64	 47.66	 95.13%   5.03E-004   95.12%   4.98E-004   
 20	3.40	-39.19	-3.40	 39.19	 96.98%	  5.02E-004   96.97%   4.97E-004
 15	1.58	-29.84	-1.58	 29.84	 98.85%   5.06E-004   98.84%   5.00E-004 
11.64	0.00	 90.00	-4.69	 4.69	100.00%	  5.00E-004
  
 
 								
Statistical strength~=1.5E-3								
gamma	 A_1	 A_2	  B_1     B_2    eta_1      S_1       eta_2      S_2
 45	11.52	-63.93	-11.52	 63.93	 94.71%   1.51E-003   94.70%   1.50E-003   
 40	10.97	-63.22	-10.97	 63.22	 94.56%	  1.50E-003   94.55%   1.49E-003
 35	9.70	-60.23	-9.70	 60.23	 94.46%   1.51E-003   94.45%   1.50E-003   
 30	7.83	-55.00	-7.83	 55.00	 95.12%   1.51E-003   95.11%   1.50E-003   
 25	5.58	-47.90	-5.58	 47.90	 96.48%	  1.50E-003   96.47%   1.49E-003
 20	3.33	-39.45	-3.33	 39.45	 98.24%	  1.50E-003   98.23%   1.49E-003
15.34	1.79	-30.61	-1.79	 30.61	100.00%	  1.50E-003
\end{verbatim}

\bibliography{prospects}

\begin{thebibliography}{41}
\expandafter\ifx\csname natexlab\endcsname\relax\def\natexlab#1{#1}\fi
\expandafter\ifx\csname bibnamefont\endcsname\relax
  \def\bibnamefont#1{#1}\fi
\expandafter\ifx\csname bibfnamefont\endcsname\relax
  \def\bibfnamefont#1{#1}\fi
\expandafter\ifx\csname citenamefont\endcsname\relax
  \def\citenamefont#1{#1}\fi
\expandafter\ifx\csname url\endcsname\relax
  \def\url#1{\texttt{#1}}\fi
\expandafter\ifx\csname urlprefix\endcsname\relax\def\urlprefix{URL }\fi
\providecommand{\bibinfo}[2]{#2}
\providecommand{\eprint}[2][]{\url{#2}}

\bibitem[{\citenamefont{Bell}(1964)}]{Bell}
\bibinfo{author}{\bibfnamefont{J.~S.} \bibnamefont{Bell}},
  \bibinfo{journal}{Physics} \textbf{\bibinfo{volume}{1}}, \bibinfo{pages}{195}
  (\bibinfo{year}{1964}).

\bibitem[{\citenamefont{Peres}(1999)}]{Peres}
\bibinfo{author}{\bibfnamefont{A.}~\bibnamefont{Peres}},
  \bibinfo{journal}{Found. Phys.} \textbf{\bibinfo{volume}{29}},
  \bibinfo{pages}{589} (\bibinfo{year}{1999}).

\bibitem[{\citenamefont{Werner and Wolf}(2001)}]{Werner}
\bibinfo{author}{\bibfnamefont{R.~F.} \bibnamefont{Werner}} \bibnamefont{and}
  \bibinfo{author}{\bibfnamefont{M.~M.} \bibnamefont{Wolf}},
  \bibinfo{journal}{Quant. Inf. Comp.} \textbf{\bibinfo{volume}{1}},
  \bibinfo{pages}{1} (\bibinfo{year}{2001}).

\bibitem[{\citenamefont{Horodecki et~al.}(2009)\citenamefont{Horodecki,
  Horodecki, Horodecki, and Horodecki}}]{Horodecki}
\bibinfo{author}{\bibfnamefont{R.}~\bibnamefont{Horodecki}},
  \bibinfo{author}{\bibfnamefont{P.}~\bibnamefont{Horodecki}},
  \bibinfo{author}{\bibfnamefont{M.}~\bibnamefont{Horodecki}},
  \bibnamefont{and}
  \bibinfo{author}{\bibfnamefont{K.}~\bibnamefont{Horodecki}},
  \bibinfo{journal}{Rev. Mod. Phys.} \textbf{\bibinfo{volume}{81}},
  \bibinfo{pages}{865} (\bibinfo{year}{2009}).

\bibitem[{\citenamefont{Clauser et~al.}(1969)\citenamefont{Clauser, Horne,
  Shimony, and Holt}}]{Clauser}
\bibinfo{author}{\bibfnamefont{J.~F.} \bibnamefont{Clauser}},
  \bibinfo{author}{\bibfnamefont{M.~A.} \bibnamefont{Horne}},
  \bibinfo{author}{\bibfnamefont{A.}~\bibnamefont{Shimony}}, \bibnamefont{and}
  \bibinfo{author}{\bibfnamefont{R.~A.} \bibnamefont{Holt}},
  \bibinfo{journal}{Phys. Rev. Lett.} \textbf{\bibinfo{volume}{23}},
  \bibinfo{pages}{880} (\bibinfo{year}{1969}).

\bibitem[{\citenamefont{Freedman and Clauser}(1972)}]{Freedman}
\bibinfo{author}{\bibfnamefont{S.~J.} \bibnamefont{Freedman}} \bibnamefont{and}
  \bibinfo{author}{\bibfnamefont{J.~F.} \bibnamefont{Clauser}},
  \bibinfo{journal}{Phys. Rev. Lett.} \textbf{\bibinfo{volume}{28}},
  \bibinfo{pages}{938} (\bibinfo{year}{1972}).

\bibitem[{\citenamefont{Genovese}(2005)}]{Genovese}
\bibinfo{author}{\bibfnamefont{M.}~\bibnamefont{Genovese}},
  \bibinfo{journal}{Phys. Rep.} \textbf{\bibinfo{volume}{413}},
  \bibinfo{pages}{319} (\bibinfo{year}{2005}).

\bibitem[{\citenamefont{Pearle}(1970)}]{Pearle}
\bibinfo{author}{\bibfnamefont{P.~M.} \bibnamefont{Pearle}},
  \bibinfo{journal}{Phys. Rev. D} \textbf{\bibinfo{volume}{2}},
  \bibinfo{pages}{1418 } (\bibinfo{year}{1970}).

\bibitem[{\citenamefont{Bell}(2004)}]{Bell2}
\bibinfo{author}{\bibfnamefont{J.~S.} \bibnamefont{Bell}},
  \emph{\bibinfo{title}{Speakable and Unspeakable in Quantum Mechanics}}
  (\bibinfo{publisher}{Cambridge University Press, Cambridge},
  \bibinfo{year}{2004}), \bibinfo{note}{pp. 139-158}.

\bibitem[{\citenamefont{Garg and Mermin}(1987)}]{Garg}
\bibinfo{author}{\bibfnamefont{A.}~\bibnamefont{Garg}} \bibnamefont{and}
  \bibinfo{author}{\bibfnamefont{N.~D.} \bibnamefont{Mermin}},
  \bibinfo{journal}{Phys. Rev. D} \textbf{\bibinfo{volume}{35}},
  \bibinfo{pages}{3831} (\bibinfo{year}{1987}).

\bibitem[{\citenamefont{Eberhard}(1993)}]{Eberhard}
\bibinfo{author}{\bibfnamefont{P.~H.} \bibnamefont{Eberhard}},
  \bibinfo{journal}{Phys. Rev. A} \textbf{\bibinfo{volume}{47}},
  \bibinfo{pages}{R747} (\bibinfo{year}{1993}).

\bibitem[{\citenamefont{Larsson and Semitecolos}(2001)}]{LarssonSemitecolos}
\bibinfo{author}{\bibfnamefont{J.-A.} \bibnamefont{Larsson}} \bibnamefont{and}
  \bibinfo{author}{\bibfnamefont{J.}~\bibnamefont{Semitecolos}},
  \bibinfo{journal}{Phys. Rev. A} \textbf{\bibinfo{volume}{63}},
  \bibinfo{pages}{022117} (\bibinfo{year}{2001}).

\bibitem[{\citenamefont{Cabello and Larsson}(2007)}]{Cabello}
\bibinfo{author}{\bibfnamefont{A.}~\bibnamefont{Cabello}} \bibnamefont{and}
  \bibinfo{author}{\bibfnamefont{J.-A.} \bibnamefont{Larsson}},
  \bibinfo{journal}{Phys. Rev. Lett.} \textbf{\bibinfo{volume}{98}},
  \bibinfo{pages}{220402} (\bibinfo{year}{2007}).

\bibitem[{\citenamefont{Brunner et~al.}(2007)\citenamefont{Brunner, Gisin,
  Scarani, and Simon}}]{Brunner}
\bibinfo{author}{\bibfnamefont{N.}~\bibnamefont{Brunner}},
  \bibinfo{author}{\bibfnamefont{N.}~\bibnamefont{Gisin}},
  \bibinfo{author}{\bibfnamefont{V.}~\bibnamefont{Scarani}}, \bibnamefont{and}
  \bibinfo{author}{\bibfnamefont{C.}~\bibnamefont{Simon}},
  \bibinfo{journal}{Phys. Rev. Lett.} \textbf{\bibinfo{volume}{98}},
  \bibinfo{pages}{220403} (\bibinfo{year}{2007}).

\bibitem[{\citenamefont{Rowe et~al.}(2001)\citenamefont{Rowe, Kielpinski,
  Meyer, Sackett, Itano, Monroe, and Wineland}}]{Rowe}
\bibinfo{author}{\bibfnamefont{M.~A.} \bibnamefont{Rowe}},
  \bibinfo{author}{\bibfnamefont{D.}~\bibnamefont{Kielpinski}},
  \bibinfo{author}{\bibfnamefont{V.}~\bibnamefont{Meyer}},
  \bibinfo{author}{\bibfnamefont{C.~A.} \bibnamefont{Sackett}},
  \bibinfo{author}{\bibfnamefont{W.~M.} \bibnamefont{Itano}},
  \bibinfo{author}{\bibfnamefont{C.}~\bibnamefont{Monroe}}, \bibnamefont{and}
  \bibinfo{author}{\bibfnamefont{D.~J.} \bibnamefont{Wineland}},
  \bibinfo{journal}{Nature} \textbf{\bibinfo{volume}{409}},
  \bibinfo{pages}{791} (\bibinfo{year}{2001}).

\bibitem[{\citenamefont{Aspect et~al.}(1982)\citenamefont{Aspect, Dalibard, and
  Roger}}]{Aspect}
\bibinfo{author}{\bibfnamefont{A.}~\bibnamefont{Aspect}},
  \bibinfo{author}{\bibfnamefont{J.}~\bibnamefont{Dalibard}}, \bibnamefont{and}
  \bibinfo{author}{\bibfnamefont{G.}~\bibnamefont{Roger}},
  \bibinfo{journal}{Phys. Rev. Lett.} \textbf{\bibinfo{volume}{49}},
  \bibinfo{pages}{1804–} (\bibinfo{year}{1982}).

\bibitem[{\citenamefont{Weihs et~al.}(1998)\citenamefont{Weihs, Jennewein,
  Simon, Weinfurter, and Zeilinger}}]{Weihs}
\bibinfo{author}{\bibfnamefont{G.}~\bibnamefont{Weihs}},
  \bibinfo{author}{\bibfnamefont{T.}~\bibnamefont{Jennewein}},
  \bibinfo{author}{\bibfnamefont{C.}~\bibnamefont{Simon}},
  \bibinfo{author}{\bibfnamefont{H.}~\bibnamefont{Weinfurter}},
  \bibnamefont{and}
  \bibinfo{author}{\bibfnamefont{A.}~\bibnamefont{Zeilinger}},
  \bibinfo{journal}{Phys. Rev. Lett.} \textbf{\bibinfo{volume}{81}},
  \bibinfo{pages}{5039} (\bibinfo{year}{1998}).

\bibitem[{\citenamefont{Tittel et~al.}(1999)\citenamefont{Tittel, Brendel,
  Gisin, and Zbinden}}]{Tittel}
\bibinfo{author}{\bibfnamefont{W.}~\bibnamefont{Tittel}},
  \bibinfo{author}{\bibfnamefont{J.}~\bibnamefont{Brendel}},
  \bibinfo{author}{\bibfnamefont{N.}~\bibnamefont{Gisin}}, \bibnamefont{and}
  \bibinfo{author}{\bibfnamefont{H.}~\bibnamefont{Zbinden}},
  \bibinfo{journal}{Phys. Rev. A} \textbf{\bibinfo{volume}{59}},
  \bibinfo{pages}{4150} (\bibinfo{year}{1999}).

\bibitem[{\citenamefont{Barrett et~al.}(2005)\citenamefont{Barrett, Hardy, and
  Kent}}]{Barrett}
\bibinfo{author}{\bibfnamefont{J.}~\bibnamefont{Barrett}},
  \bibinfo{author}{\bibfnamefont{L.}~\bibnamefont{Hardy}}, \bibnamefont{and}
  \bibinfo{author}{\bibfnamefont{A.}~\bibnamefont{Kent}},
  \bibinfo{journal}{Phys. Rev. Lett.} \textbf{\bibinfo{volume}{95}},
  \bibinfo{pages}{010503} (\bibinfo{year}{2005}).

\bibitem[{\citenamefont{Masanes et~al.}(2009)\citenamefont{Masanes, Renner,
  Winter, Barrett, and Christandl}}]{Masanes1}
\bibinfo{author}{\bibfnamefont{L.}~\bibnamefont{Masanes}},
  \bibinfo{author}{\bibfnamefont{R.}~\bibnamefont{Renner}},
  \bibinfo{author}{\bibfnamefont{A.}~\bibnamefont{Winter}},
  \bibinfo{author}{\bibfnamefont{J.}~\bibnamefont{Barrett}}, \bibnamefont{and}
  \bibinfo{author}{\bibfnamefont{M.}~\bibnamefont{Christandl}}
  (\bibinfo{year}{2009}), \eprint{arXiv:quant-ph/0606049v4}.

\bibitem[{\citenamefont{Masanes}(2009)}]{Masanes2}
\bibinfo{author}{\bibfnamefont{L.}~\bibnamefont{Masanes}},
  \bibinfo{journal}{Phys. Rev. Lett.} \textbf{\bibinfo{volume}{102}},
  \bibinfo{pages}{140501} (\bibinfo{year}{2009}).

\bibitem[{\citenamefont{Lita et~al.}(2008)\citenamefont{Lita, Miller, and
  Nam}}]{Lita}
\bibinfo{author}{\bibfnamefont{A.~E.} \bibnamefont{Lita}},
  \bibinfo{author}{\bibfnamefont{A.~J.} \bibnamefont{Miller}},
  \bibnamefont{and} \bibinfo{author}{\bibfnamefont{S.~W.} \bibnamefont{Nam}},
  \bibinfo{journal}{Opt. Express} \textbf{\bibinfo{volume}{16}},
  \bibinfo{pages}{3032} (\bibinfo{year}{2008}).

\bibitem[{\citenamefont{Shih and Alley}(1988)}]{Shih}
\bibinfo{author}{\bibfnamefont{Y.~H.} \bibnamefont{Shih}} \bibnamefont{and}
  \bibinfo{author}{\bibfnamefont{C.~O.} \bibnamefont{Alley}},
  \bibinfo{journal}{Phys. Rev. Lett.} \textbf{\bibinfo{volume}{61}},
  \bibinfo{pages}{2921} (\bibinfo{year}{1988}).

\bibitem[{\citenamefont{Ou and Mandel}(1988)}]{Ou}
\bibinfo{author}{\bibfnamefont{Z.~Y.} \bibnamefont{Ou}} \bibnamefont{and}
  \bibinfo{author}{\bibfnamefont{L.}~\bibnamefont{Mandel}},
  \bibinfo{journal}{Phys. Rev. Lett.} \textbf{\bibinfo{volume}{61}},
  \bibinfo{pages}{50} (\bibinfo{year}{1988}).

\bibitem[{\citenamefont{Kiess et~al.}(1993)\citenamefont{Kiess, Shih,
  Sergienko, and Alley}}]{Kiess}
\bibinfo{author}{\bibfnamefont{T.~E.} \bibnamefont{Kiess}},
  \bibinfo{author}{\bibfnamefont{Y.~H.} \bibnamefont{Shih}},
  \bibinfo{author}{\bibfnamefont{A.~V.} \bibnamefont{Sergienko}},
  \bibnamefont{and} \bibinfo{author}{\bibfnamefont{C.~O.} \bibnamefont{Alley}},
  \bibinfo{journal}{Phys. Rev. Lett.} \textbf{\bibinfo{volume}{71}},
  \bibinfo{pages}{3893} (\bibinfo{year}{1993}).

\bibitem[{\citenamefont{Lounis and Orrit}(2005)}]{Lounis}
\bibinfo{author}{\bibfnamefont{B.}~\bibnamefont{Lounis}} \bibnamefont{and}
  \bibinfo{author}{\bibfnamefont{M.}~\bibnamefont{Orrit}},
  \bibinfo{journal}{Rep. Prog. Phys.} \textbf{\bibinfo{volume}{68}},
  \bibinfo{pages}{1129} (\bibinfo{year}{2005}).

\bibitem[{\citenamefont{Oxborrow and Sinclair}(2005)}]{Oxborrow}
\bibinfo{author}{\bibfnamefont{M.}~\bibnamefont{Oxborrow}} \bibnamefont{and}
  \bibinfo{author}{\bibfnamefont{A.~G.} \bibnamefont{Sinclair}},
  \bibinfo{journal}{Contemp. Phys.} \textbf{\bibinfo{volume}{46}},
  \bibinfo{pages}{173} (\bibinfo{year}{2005}).

\bibitem[{\citenamefont{van Dam et~al.}(2005)\citenamefont{van Dam, Gill, and
  Grunwald}}]{VanDam}
\bibinfo{author}{\bibfnamefont{W.}~\bibnamefont{van Dam}},
  \bibinfo{author}{\bibfnamefont{R.~D.} \bibnamefont{Gill}}, \bibnamefont{and}
  \bibinfo{author}{\bibfnamefont{P.~D.} \bibnamefont{Grunwald}},
  \bibinfo{journal}{IEEE Trans. Inf. Theory} \textbf{\bibinfo{volume}{51}},
  \bibinfo{pages}{2812} (\bibinfo{year}{2005}).

\bibitem[{\citenamefont{Kwiat et~al.}(1994)\citenamefont{Kwiat, Eberhard,
  Steinberg, and Chiao}}]{Kwiat}
\bibinfo{author}{\bibfnamefont{P.~G.} \bibnamefont{Kwiat}},
  \bibinfo{author}{\bibfnamefont{P.~H.} \bibnamefont{Eberhard}},
  \bibinfo{author}{\bibfnamefont{A.~M.} \bibnamefont{Steinberg}},
  \bibnamefont{and} \bibinfo{author}{\bibfnamefont{R.~Y.} \bibnamefont{Chiao}},
  \bibinfo{journal}{Phys. Rev. A} \textbf{\bibinfo{volume}{49}},
  \bibinfo{pages}{3209} (\bibinfo{year}{1994}).

\bibitem[{\citenamefont{DeCaro and Garuccio}(1994)}]{Caro}
\bibinfo{author}{\bibfnamefont{L.}~\bibnamefont{DeCaro}} \bibnamefont{and}
  \bibinfo{author}{\bibfnamefont{A.}~\bibnamefont{Garuccio}},
  \bibinfo{journal}{Phys. Rev. A} \textbf{\bibinfo{volume}{50}},
  \bibinfo{pages}{R2803} (\bibinfo{year}{1994}).

\bibitem[{\citenamefont{Popescu et~al.}(1997)\citenamefont{Popescu, Hardy, and
  Zukowski}}]{Popescu}
\bibinfo{author}{\bibfnamefont{S.}~\bibnamefont{Popescu}},
  \bibinfo{author}{\bibfnamefont{L.}~\bibnamefont{Hardy}}, \bibnamefont{and}
  \bibinfo{author}{\bibfnamefont{M.}~\bibnamefont{Zukowski}},
  \bibinfo{journal}{Phys. Rev. A} \textbf{\bibinfo{volume}{56}},
  \bibinfo{pages}{R4353} (\bibinfo{year}{1997}).

\bibitem[{\citenamefont{Bahadur}(1967)}]{Bahadur}
\bibinfo{author}{\bibfnamefont{R.~R.} \bibnamefont{Bahadur}}, in
  \emph{\bibinfo{booktitle}{Proc. Fifth Berkeley Symp. on Math. Statist. and
  Prob.}} (\bibinfo{publisher}{Univ. of Calif. Press}, \bibinfo{year}{1967}),
  vol.~\bibinfo{volume}{1}, pp. \bibinfo{pages}{13--26}.

\bibitem[{\citenamefont{Vardi and Lee}(1993)}]{Vardi}
\bibinfo{author}{\bibfnamefont{Y.}~\bibnamefont{Vardi}} \bibnamefont{and}
  \bibinfo{author}{\bibfnamefont{D.}~\bibnamefont{Lee}}, \bibinfo{journal}{J.
  Royal Stat. Soc. B} \textbf{\bibinfo{volume}{55}}, \bibinfo{pages}{569}
  (\bibinfo{year}{1993}).

\bibitem[{\citenamefont{Vertesi et~al.}(2009)\citenamefont{Vertesi, Pironio,
  and Brunner}}]{Vertesi}
\bibinfo{author}{\bibfnamefont{T.}~\bibnamefont{Vertesi}},
  \bibinfo{author}{\bibfnamefont{S.}~\bibnamefont{Pironio}}, \bibnamefont{and}
  \bibinfo{author}{\bibfnamefont{N.}~\bibnamefont{Brunner}}
  (\bibinfo{year}{2009}), \eprint{arXiv:0909.3171v2 [quant-ph]}.

\bibitem[{\citenamefont{White et~al.}(1999)\citenamefont{White, James,
  Eberhard, and Kwiat}}]{White}
\bibinfo{author}{\bibfnamefont{A.~G.} \bibnamefont{White}},
  \bibinfo{author}{\bibfnamefont{D.~F.~V.} \bibnamefont{James}},
  \bibinfo{author}{\bibfnamefont{P.~H.} \bibnamefont{Eberhard}},
  \bibnamefont{and} \bibinfo{author}{\bibfnamefont{P.~G.} \bibnamefont{Kwiat}},
  \bibinfo{journal}{Phys. Rev. Lett.} \textbf{\bibinfo{volume}{83}},
  \bibinfo{pages}{3103} (\bibinfo{year}{1999}).

\bibitem[{\citenamefont{Brida et~al.}(2000)\citenamefont{Brida, Genovese,
  Novero, and Predazzi}}]{Brida}
\bibinfo{author}{\bibfnamefont{G.}~\bibnamefont{Brida}},
  \bibinfo{author}{\bibfnamefont{M.}~\bibnamefont{Genovese}},
  \bibinfo{author}{\bibfnamefont{C.}~\bibnamefont{Novero}}, \bibnamefont{and}
  \bibinfo{author}{\bibfnamefont{E.}~\bibnamefont{Predazzi}},
  \bibinfo{journal}{Phys. Lett. A} \textbf{\bibinfo{volume}{268}},
  \bibinfo{pages}{12} (\bibinfo{year}{2000}).

\bibitem[{\citenamefont{Gisin}(1991)}]{Gisin}
\bibinfo{author}{\bibfnamefont{N.}~\bibnamefont{Gisin}},
  \bibinfo{journal}{Phys. Lett. A} \textbf{\bibinfo{volume}{154}},
  \bibinfo{pages}{201} (\bibinfo{year}{1991}).

\bibitem[{\citenamefont{Popescu and Rohrlich}(1992)}]{Popescu2}
\bibinfo{author}{\bibfnamefont{S.}~\bibnamefont{Popescu}} \bibnamefont{and}
  \bibinfo{author}{\bibfnamefont{D.}~\bibnamefont{Rohrlich}},
  \bibinfo{journal}{Phys. Lett. A} \textbf{\bibinfo{volume}{166}},
  \bibinfo{pages}{293} (\bibinfo{year}{1992}).

\bibitem[{\citenamefont{Scarani and Gisin}(2001)}]{Scarani}
\bibinfo{author}{\bibfnamefont{V.}~\bibnamefont{Scarani}} \bibnamefont{and}
  \bibinfo{author}{\bibfnamefont{N.}~\bibnamefont{Gisin}}, \bibinfo{journal}{J.
  Phys. A: Math. Gen.} \textbf{\bibinfo{volume}{34}}, \bibinfo{pages}{6043}
  (\bibinfo{year}{2001}).

\bibitem[{\citenamefont{Wasilewski et~al.}(2006)\citenamefont{Wasilewski,
  Lvovsky, Banaszek, and Radzewicz}}]{Wasilewski}
\bibinfo{author}{\bibfnamefont{W.}~\bibnamefont{Wasilewski}},
  \bibinfo{author}{\bibfnamefont{A.~I.} \bibnamefont{Lvovsky}},
  \bibinfo{author}{\bibfnamefont{K.}~\bibnamefont{Banaszek}}, \bibnamefont{and}
  \bibinfo{author}{\bibfnamefont{C.}~\bibnamefont{Radzewicz}},
  \bibinfo{journal}{Phys. Rev. A} \textbf{\bibinfo{volume}{73}},
  \bibinfo{pages}{063819} (\bibinfo{year}{2006}).

\bibitem[{\citenamefont{Lvovsky et~al.}(2007)\citenamefont{Lvovsky, Wasilewski,
  and Banaszek}}]{Lvovsky}
\bibinfo{author}{\bibfnamefont{A.~I.} \bibnamefont{Lvovsky}},
  \bibinfo{author}{\bibfnamefont{W.}~\bibnamefont{Wasilewski}},
  \bibnamefont{and} \bibinfo{author}{\bibfnamefont{K.}~\bibnamefont{Banaszek}},
  \bibinfo{journal}{J. Mod. Optics} \textbf{\bibinfo{volume}{54}},
  \bibinfo{pages}{721} (\bibinfo{year}{2007}).

\end{thebibliography}
\end{document}